\documentstyle[12pt,epsf]{article}
\advance \textheight by \topskip
\begin{document}
\begin{titlepage}
\hbox to \hsize{{\hfil}}
\hbox to \hsize{{\hfil }}
\hbox to \hsize{{\hfil }}
\vfill
\large \bf
\begin{center}
  LOW-$x$ BEHAVIOR OF SPIN STRUCTURE FUNCTIONS: UNITARITY 
AND NONPERTURBATIVE HADRON STRUCTURE
  
\end{center}
\vskip 1cm
\normalsize
\begin{center}
{\bf S. M. Troshin and N. E. Tyurin}\\[1ex]
{\small  \it Institute for High Energy Physics,\\
Protvino, Moscow Region, 142284 Russia}
\end{center}
\vskip 1.5cm
\begin{abstract}
We consider low-$x$ behavior
of the spin structure functions
$g_1(x)$ and $h_1(x)$ in the unitarized chiral quark model that 
combines the ideas on  constituent quark structure of hadrons with
a geometrical scattering picture and unitarity.  A nondiffractive
singular low-$x$  dependence of $g^p_1(x)$ and $g_1^n(x)$ 
is obtained and a diffractive type
smooth behavior of $h_1(x)$  is  predicted at small $x$.
\end{abstract}
\vfill
\end{titlepage}

Experimental evaluation
of the first moments of $g_1$ and $h_1$ (and
 the total nucleon helicity carried
by  quarks and tensor charge respectively) in principle
are sensitive to
a particular theoretical
extrapolation of the structure functions $g_1(x)$ and $h_1(x)$ to $x=0$.
The  essential point in the study of low-$x$ dynamics
is that the space-time structure of
the scattering
at small values of $x$ involves  large distances
$l\sim 1/Mx$ on the light--cone \cite{pas} and the
region $x\sim 0$ is therefore determined by the 
nonperturbative dynamics.
A number of models attributes the observed increase of
 $g_1(x)$ at small $x$ to
the diffractive contribution. Such contribution being dominant at 
 smallest values of $x$ would lead to the ``equal''
 structure functions $g_1^p(x)$ and $g_1^n(x)$ in this kinematical 
 region, i. e.
 \[
 g_1^p(x)/g_1^n(x)\to 1
 \]
 at $x\to 0$.
 Such behavior has not been confirmed in the recent experiments. In particular,
 the  SMC data \cite{smc} demonstrate the following approximate relation
  in the region of
$0.003\leq x \leq 0.1$:
\[
g_1^p(x)\simeq-g_1^n(x).
\]

To consider low-$x$ region and
obtain the explicit forms for the quark spin densities
 $\Delta q(x)$ and $\delta q(x)$ at $x\to 0$ it is convenient to use the relations
 between these functions and  discontinuities of the helicity
amplitudes of the  antiquark--hadron forward
scattering \cite{jaffesof}. We use a nonperturbative approach where
 unitarity is explicitly taken into account
 via unitary representations for the helicity amplitudes, 
which follow
from their relations  to the $U$--matrix \cite{abk}

In the model a quark is considered as a structured hadronlike object since at small
$x$ the photon converts to a quark pair at  long distance before
it interacts with the hadron. At large distances perturbative QCD
vacuum  undergoes  transition into a nonperturbative one with
formation of the  quark condensate. Appearance of the condensate
means the spontaneous   chiral symmetry breaking and the current quark
transforms into a massive quasiparticle state -- a constituent quark.
Constituent quark is embedded into the nonperturbative vacuum
(condensate)  and therefore we can 
treat it similar to a  hadron.
  Spin of constituent quark $J_{U}$  in this approach is given
 by the  following sum \[
 J_{U}=1/2  =  S_{u_v}+S_{\{\bar q q\}}+\langle L_{\{\bar qq\}}\rangle=
               1/2+S_{\{\bar q q\}}+\langle L_{\{\bar qq\}}\rangle.
\]
It is also important to note the exact compensation between the spins
of quark--antiquark pairs  and their angular orbital momenta, i.e.
$\langle L_{\{\bar q q\}}\rangle= -S_{\{\bar q q\}}$.
 
We consider effective lagrangian approach where gluon
 degrees of freedom are overintegrated.
The value of the orbital
 momentum contribution into the spin of constituent quark can be
 estimated according to  the relation between
 contributions of current quarks into a proton spin and corresponding
contributions of current quarks into a spin of the constituent quarks
and that of the constituent quarks into  the proton spin.
 The existence of this orbital angular momentum, i.e.
 orbital motion of quark matter inside constituent quark, is the
 origin of the observed asymmetries in inclusive production at
  moderate and high transverse momenta.
Mechanism of quark helicity flip in this picture is associated with
the constituent quark interaction with the quark generated under
interaction of the condensates \cite{abk}.
 Quark exchange process between the valence quark and an
appropriate quark with relevant orientation
of its spin and the same flavor will provide the necessary helicity
flip transition, i.e. $Q_+\rightarrow Q_-$.

The helicity amplitudes $F_{1,2,3}(s,t)|_{t=0}$ 
at high values of $s$ and  then  the  functional
dependencies for the quark densities
$q(x)$, $\Delta q(x)$ and $\delta q(x)$ at small $x$ were
obtained \cite{g1h1}.

The low-$x$ behavior of quark spin densities is as follows:
\[
q(x)\sim \frac{1}{x}\ln^2(1/x),\quad \Delta q(x)\sim \frac{1}{\sqrt{x}}\ln(1/x),\quad \delta q(x)\sim x^{c}\ln(1/x),
\]
and correspondingly
\[
F^p_1(x)/F_1^n(x)\to 1,\quad h^p_1(x)/h_1^n(x)\to 1
\]
at $x \to 0$, with the explicit forms  as follows
\[
F^p_1(x)\sim
\frac{1}{x} \ln^2(1/x),\quad
h^p_1(x)\sim
 x^{c}\ln(1/x).
\]

 Comparison of the spin structure function $g_1(x)$
 with the SMC data provides  
a satisfactory agreement 
 with experiment at small $x$
($0<x<0.1$)  and  leads to the values $C^p=2.07\cdot 10^{-2}$ and 
$C^n=-2.10\cdot 10^{-2}$ (cf. Fig. 1).
 \begin{figure}[t]
 \hspace*{3cm}
 \epsfxsize=80  mm  \epsfbox{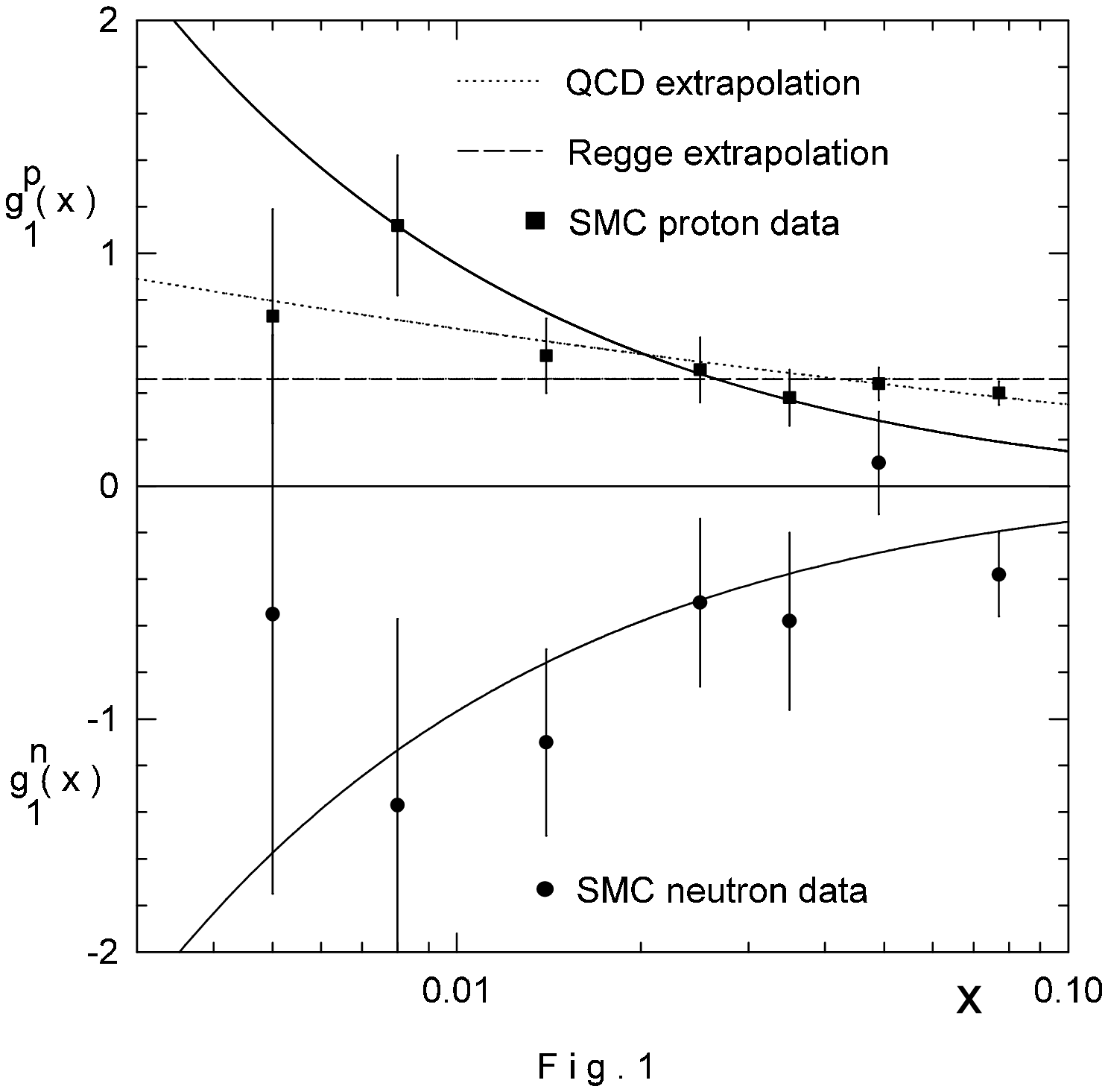}
 
 %\caption{Low-$x$ behavior of the spin structure functions
 %$g_1^p(x)$ and $g_1^n(x)$}
 
 \end{figure}

 %\begin{figure}[t]
%\epsfxsize= 60 mm \epsfbox{fig1n.eps}
% \caption{Low-$x$ behavior of the spin structure functions
% $g_1^p(x)$ and $g_1^n(x)$}  
%\end{figure}
 The functional dependence of the spin structure
functions
\[
g_1^{p,n}(x)\sim\frac{1}{\sqrt{x}}\ln(1/x)
\]
is in a good agreement with the new E154, E155 \cite{nlo}
and HERMES data
\cite{herm} as well. The model leads to the approximate relation
\[
g^p_1(x)/g_1^n(x)\simeq -1
\]
at small values of $x$.

The above extrapolation of $g_1(x)$ at small $x$
 provides the following
approximate values for the quark spin contributions:
\[
\Delta\Sigma  \simeq  0.25,\quad
\Delta u  \simeq  0.81,\quad
\Delta d  \simeq  -0.45,\quad
\Delta s  \simeq  -0.11, \label{sp}
\]
which demonstrate that the singular behavior
of $g_1^p(x)$  does not lead to significant
deviations from the results of the experimental analysis \cite{smc}
where the smooth extrapolation of the data to $x=0$ was used.

\end{document}